\documentstyle[psfig]{mn}

\newcommand{\bls}{BL Lac objects}
\newcommand{\mkn}{Mrk\,421}
\newcommand{\lc}{light curve}
\newcommand{\lcs}{light curves}

\newcommand{\ts}{timescale}
\newcommand{\tss}{timescales}
\newcommand{\tcool}{$t_{\rm cool}$}
\newcommand{\tacc}{$t_{\rm acc}$}
\newcommand{\sy}{synchrotron}

\newcommand{\sax}{{\it Beppo}SAX}
\newcommand{\etal}{et al.\,}
\newcommand{\beq}{\begin{equation}}
\newcommand{\eeq}{\end{equation}}

\begin{document}

\title[Cross-spectral analysis of the X-ray variability of Mrk~421]
{Cross-spectral analysis of the X-ray variability of Markarian~421}

\author[Y.H. Zhang] 
{Y.H. Zhang \thanks{E-mail: youhong.zhang@uninsubria.it} \\
	Dipartimento di Scienze, Universit\`a dell'Insubria,
	via Valleggio 11, I-22100 Como, Italy}
% \\
%{\tt e-mail: youhong.zhang@uninsubria.it} }

\date{Accepted .
	Received ;
      in original form }

\pagerange{\pageref{firstpage}--\pageref{lastpage}}
\pubyear{2002} 

\maketitle

\label{firstpage}

\begin{abstract}
Using the cross-spectral method, we confirm the existence of the X-ray 
hard lags discovered with cross-correlation function technique during a large
flare of \mkn\ observed with \sax . For the 0.1--2 versus 2--10~keV light 
curves, both methods suggest sub-hour hard lags. In the time domain, 
the degree of hard lag, i.e., the amplitude of the 3.2--10~keV photons 
lagging the lower energy ones, tends to increase with the decreasing energy. 
In the Fourier frequency domain, by investigating the cross-spectra of the  
0.1--2/2--10~keV and the 2--3.2/3.2--10~keV pairs of light curves, the 
flare also shows hard lags at the lowest frequencies. However, with the 
present data, it is impossible to constrain the dependence of the lags 
on frequencies even though the detailed simulations demonstrate that the 
hard lags at the lowest frequencies probed by the flare are not an 
artifact of sparse sampling, Poisson and red noise. 
As a possible interpretation, the implication of the hard lags is 
discussed in the context of the interplay between the (diffusive) 
acceleration and \sy\ cooling of relativistic electrons responsible for 
the observed X-ray emission. The energy-dependent hard lags are in 
agreement with the expectation of an energy-dependent acceleration 
timescale. The inferred magnetic field ($B \sim 0.11$~Gauss) is consistent 
with the value inferred from the Spectral Energy Distributions of the source. 
Future investigations with higher quality data that whether or not the time 
lags are energy-/frequency-dependent will provide a new constraint on 
the current models of the TeV blazars. 
\end{abstract}

		%--------------------%

\section{Introduction}\label{sec:intro}

It has been well established that blazars are extra-galactic sources
possessing relativistic jets aligned close to the line of sight,
nevertheless, it is still poorly understood how the jets are
powered, formed and collimated, and how particles are efficiently
accelerated. One of the best observational approaches would be to use 
temporal and spectral analysis of the emission from the
jets. Although blazars are not the unique hosts of jets, being
dominated by Doppler effects (causing the observed emission to be 
enhanced and the \tss\ shortened), they are the ideal targets for
studying jet physics.  

The dominant radiation mechanisms in blazars are thought to be
\sy\ and inverse Compton by relativistic electrons in a 
tangled magnetic field, which can reproduce the two peaks of
the Spectral Energy Distributions (SEDs) in the $\nu F_{\nu} -
\nu$ diagram: synchrotron radiation is responsible for the low energy
peak, while inverse-Compton upscattering by the same population of
electrons produces the high energy one (e.g., Urry \& Padovani
1995). In such a picture, the blazar family could be unified on the basis
of the SEDs whose properties are determined by the bolometric luminosities 
(Ghisellini \etal 1998).

According to this scenario, the variability of blazars is expected to be
energy-dependent, with the highest energy part of each emission
component showing the most rapid variations as produced by the highest
energy part of the relativistic electron distribution which evolves
most rapidly. For high-energy (usually UV/soft X-rays) \sy\
peaked blazars (HBLs), X-rays provide an ideal radiative
window for studying the variations because (1) rapid variability
indicates that the X-rays arise from the innermost region of the jets,
and give direct clues on the central source; (2) \sy\ X-ray emission
probes the electrons accelerated to the highest energies, which plausibly
have the longest acceleration and the shortest cooling times. 

The detected TeV blazars at TeV energies are typical HBLs,  
including three well-studied classical \bls , Mrk~421, Mrk~501,
and PKS~2155--304. The former two have been detected as bright and 
variable TeV emitters. In particular, \mkn\ ($z=0.031$) is the brightest
blazar at UV, X-ray and TeV wavelengths. These sources have thus received
particular attention as ideal targets for detailed temporal and spectral
variability studies in the broadest spectral ranges. Extensive
multi-wavelength monitoring campaigns and long looks with various
satellites have been conducted.

Intensive monitoring has shown that the synchrotron peak energy
up-shifts with flux in these sources (Pian \etal 1998; Fossati \etal
2000b; Tavecchio \etal 2001; Zhang \etal 2002). They have also 
exhibited quite different variability properties and spectral
evolution from flare to flare, indicating that the high energy photons 
can lead or lag the low energy ones (e.g., Zhang \etal 1999, 2002; 
Fossati \etal 2000a; Tanihata \etal 2001). The so-called soft lag 
(lower energy X-ray photons lagging higher energy ones) is 
consistent with the picture of the energy-dependent cooling time of 
relativistic electrons---higher energy electrons cool faster. The 
opposite behavior, i.e., the so-called hard lag (higher energy X-ray
photons lagging the lower energy ones) has been found (as not
rare) with recent long looks of the three TeV blazars with \sax\ and
ASCA. This ``unusual'' hard lag has been thought to give direct
information on electrons acceleration: it takes longer time for higher
energy electrons to accelerate to the radiative energy.

In this paper, through the studies of the time lags in the time and 
frequency domains, the cross-correlation function (CCF) and the 
cross-spectral methods are used to re-examine the discovery of the X-ray 
hard lags during a large flare of \mkn\ detected by \sax\
(Fossati \etal 2000a). The latter method was historically used to analyze
the X-ray variability of Galactic black hole candidates (GBHCs; e.g., 
Miyamoto \etal 1988, 1991; Nowak \etal 1999). Recently, Papadakis, Nandra
\& Kazanas (2001) adopted this technique to study the X-ray variability
of a Seyfert galaxy. Since the cross-spectrum can give more information 
(i.e., the Fourier frequency-dependent time lags) than the CCF can do, 
it is able to impose stronger constraints on the emission models.  

The paper is organized as follows. The \lcs\ of the flare are presented
in \S\ref{sec:lc}. The characteristic feature of the hard lag is
examined by showing the evolutionary behavior of the hardness
ratio versus the count rate. In \S\ref{sec:lage} the dependence of
hard lags on photon energies is studied with the CCF method incorporating 
with a model-independent Monte Carlo simulations. 
We investigate in \S\ref{sec:flag:results} the time lags in Fourier 
frequency domain using the cross-spectral technique; detailed simulations 
are performed in \S\ref{sec:flag:simulations} to investigate the effects 
of Poisson and red noise, sampling and signal-noise (S/N) ratio of 
the data sets. In \S\ref{sec:disc:summary} we discuss and compare the 
results derived in time and frequency domains; the physical implications 
of the results are preliminarily explored in \S\ref{sec:disc:implications}; 
we also briefly compare in \S\ref{sec:disc:comparison} the time lags in 
different black hole accreting systems. Finally, we present our conclusions 
in \S\ref{sec:conc}. 

		%--------------------%

\section{Light Curves and Hardness Ratios} \label{sec:lc}

%-----------------------------
\begin{figure*}
\begin{tabular}{cc}
\psfig{file=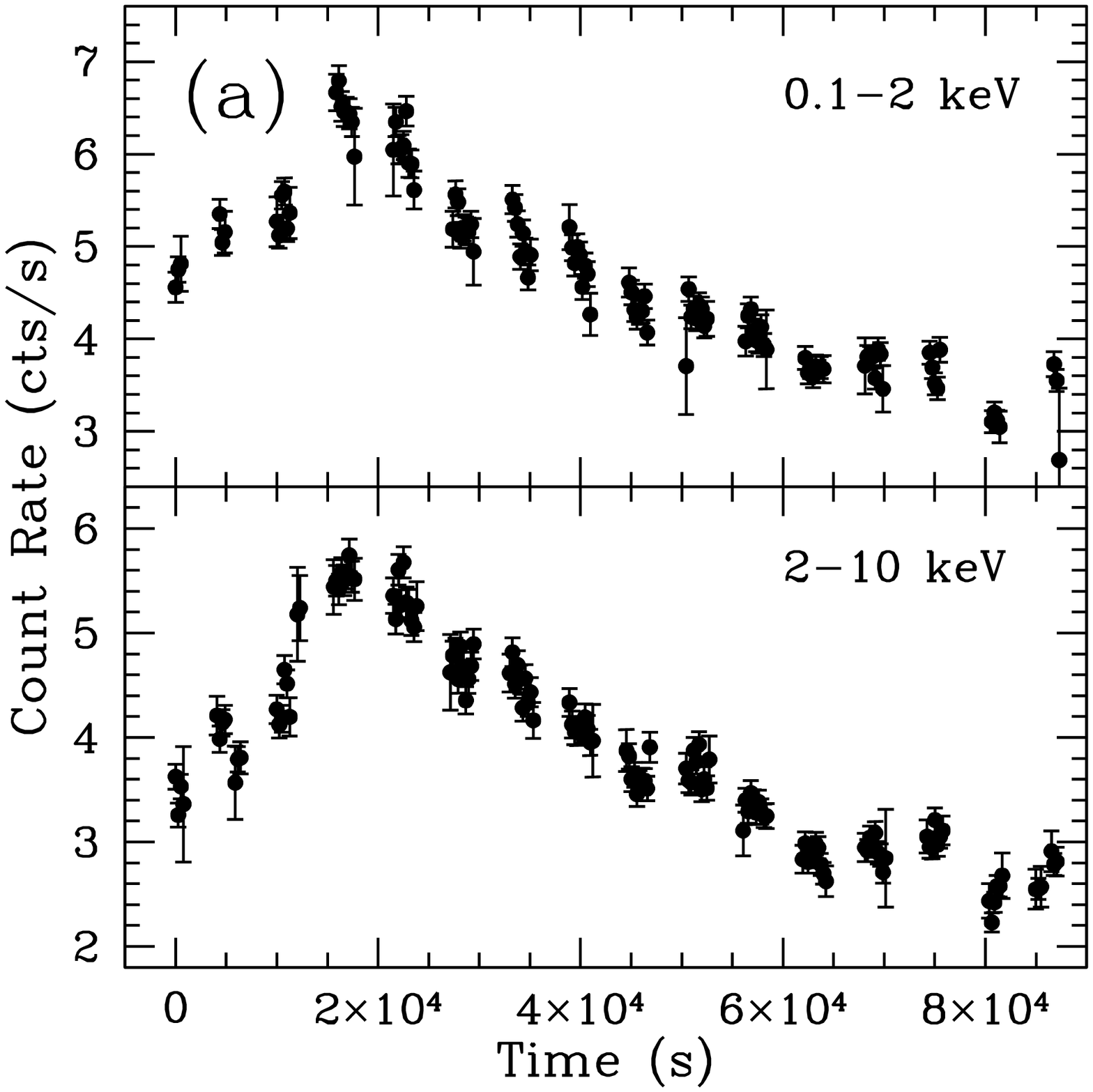,width=8cm,height=8cm} \hspace{8mm} 
\psfig{file=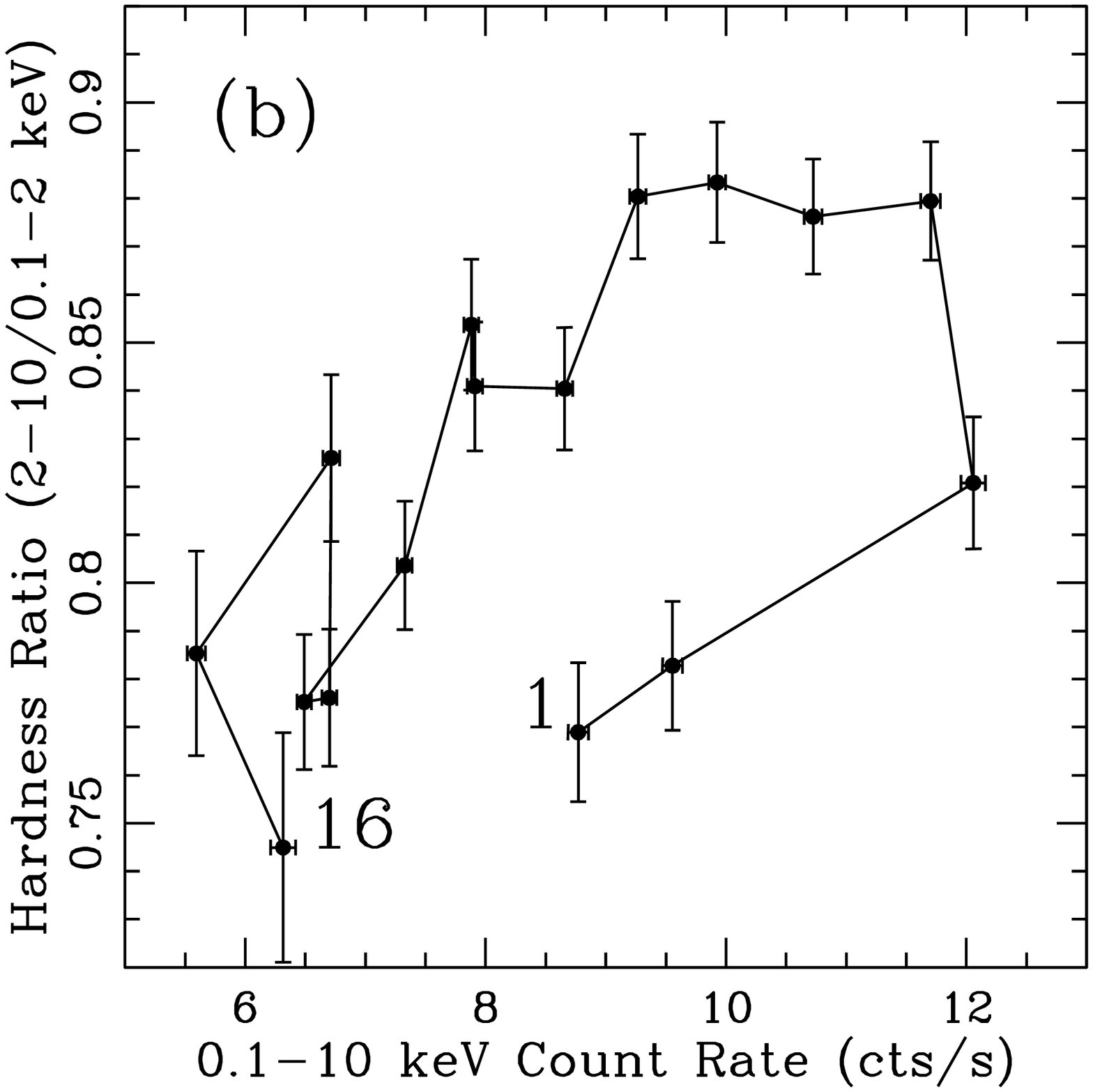,width=8cm,height=8cm}
\end{tabular}
\caption{ 
(a) Light curves of the flare of \mkn\ detected with \sax\
on 21 April 1998. The data are binned over 256~s. The rising phase
of the flare is possibly incomplete.
(b) Hardness ratios of the 2--10~keV to the 0.1--2~keV bands as a
function of the observed count rates in the 0.1--10~keV band. The data are
binned over the \sax\ orbital period. The first and last point are
numbered with the orbital number 1 and 16, respectively, and the
evolutionary direction follows the connected line from 1 to 16. An
anticlockwise loop can be seen clearly, suggesting that the 2--10~keV 
emission lags the 0.1--2~keV one. }
\label{fig:lc}
\end{figure*}
%-------------------------------------------

\sax\ (Boella \etal 1997 and references therein) observed \mkn\ on
21--23 April 1998 as part of a multi-wavelength long monitoring
campaign involving \sax , ASCA, RXTE, EUVE, and ground-based TeV
observatories (Maraschi \etal 1999; Takahashi \etal 2000). Full details of
\sax\ data reduction have been given in Fossati \etal (2000a). The \sax\
observations consist of two distinct parts. This work concentrates on the 
large flare detected on 21 April, of which the \lc\ are shown in
Figure~\ref{fig:lc}a in two energy bands, i.e., 0.1--2~keV (LECS) and
2--10~keV (MECS). For the \lcs\ of the second part of the observation we
refer to Fossati \etal (2000a). 
 
Time-resolved spectral analysis for the flare has been performed 
by Fossati \etal (2000b). One of the interesting results relevant
to this work is that the evolution of the flare in the plane of 
spectral index versus flux follows an anticlockwise loop (see
their Figure~5), indicating the presence of an X-ray hard lag. This 
qualitatively confirmed the discovery of clear hard lag by the 
CCF techniques (Fossati \etal 2000a).

Time-resolved spectral analysis can be simply performed by considering  
hardness ratio, which is a simple representation of the two-point
spectral index. For this reason, we show in Figure~\ref{fig:lc}b 
the hardness ratio of 2--10 to 0.1--2~keV versus the 0.1--10~keV count 
rate, which is binned over the \sax\ orbital period ($\sim 5670$~s) --- 
the minimum time bin size over which the \lcs\ can be evenly sampled
without differentiating exposure efficiency between the LECS and the
MECS detectors. It is clear from Figure~\ref{fig:lc}b that 
the evolution of the flare tracks a well-defined anticlockwise 
direction, fully consistent with the discovery of the time-resolved
spectral analysis. 
     
		%--------------------%

\section{Energy-dependence of Time Lags} \label{sec:lage}

To quantify the hard lags between the low and high energy
variations, we calculated the CCF with two techniques suited to unevenly 
sampled time series: the
Discrete Correlation Function (DCF, Edelson \& Krolik 1988) and
Modified Mean Deviation (MMD, Hufnagel \& Bregman 1992), incorporating 
with a model-independent Monte Carlo simulations taking into
account ``flux redistribution'' (FR) and ``random subset selection''
(RSS) of the two cross-correlated light curves (Peterson \etal 1998) to 
statistically determine the significance of any time lags from the 
cross-correlation peak distribution (CCPD; Maoz \& Netzer 1989). Such an 
procedure suggests that the 3.5--10~keV emission lags the 0.1--1.5~keV one 
by $2.7^{+1.9}_{-1.2}$ (DCF) and $2.3^{+1.2}_{-0.7}$ (MMD) 
($68\%$ confidence range with respect to the average of the CCPD; 
Fossati \etal 2000a).

Here we re-estimate the lag using the \textit{Fisher's z-transformed} DCF 
(ZDCF; Alexander 1997) method that approximates each DCF point as a normal 
distribution. We also take into account some more accurate issues: 
(1) the ZDCF is normalized by the mean and standard deviation of 
the two cross-correlated light curves using only the data points that 
actually contribute to the calculation of each lag (White \& Peterson 1994), 
i.e., the so-called ``local'' CCF (Welsh 1999). With respect to the 
``standard'' CCF that is identically normalized by the mean and standard 
deviation of the whole time series, the ``local'' CCF definitely avoids the 
problem that the DCF amplitudes can be significantly smaller than $-1$ at some 
lags, which is commonly seen in the literature (e.g., Tanihata \etal 2001). 
We also noticed such problem from the data sets studied here, and from 
other \sax\ observations of Mrk~421 and PKS~2155--304, and in particular 
from an XMM-Newton observation of PKS~2155--304 (Maraschi \etal 2002); 
(2) the ZDCF is binned by equal points rather than by equal lag step.

%----------------------------------------
\begin{figure*}
\begin{tabular}{cc}
\centerline{
\psfig{file=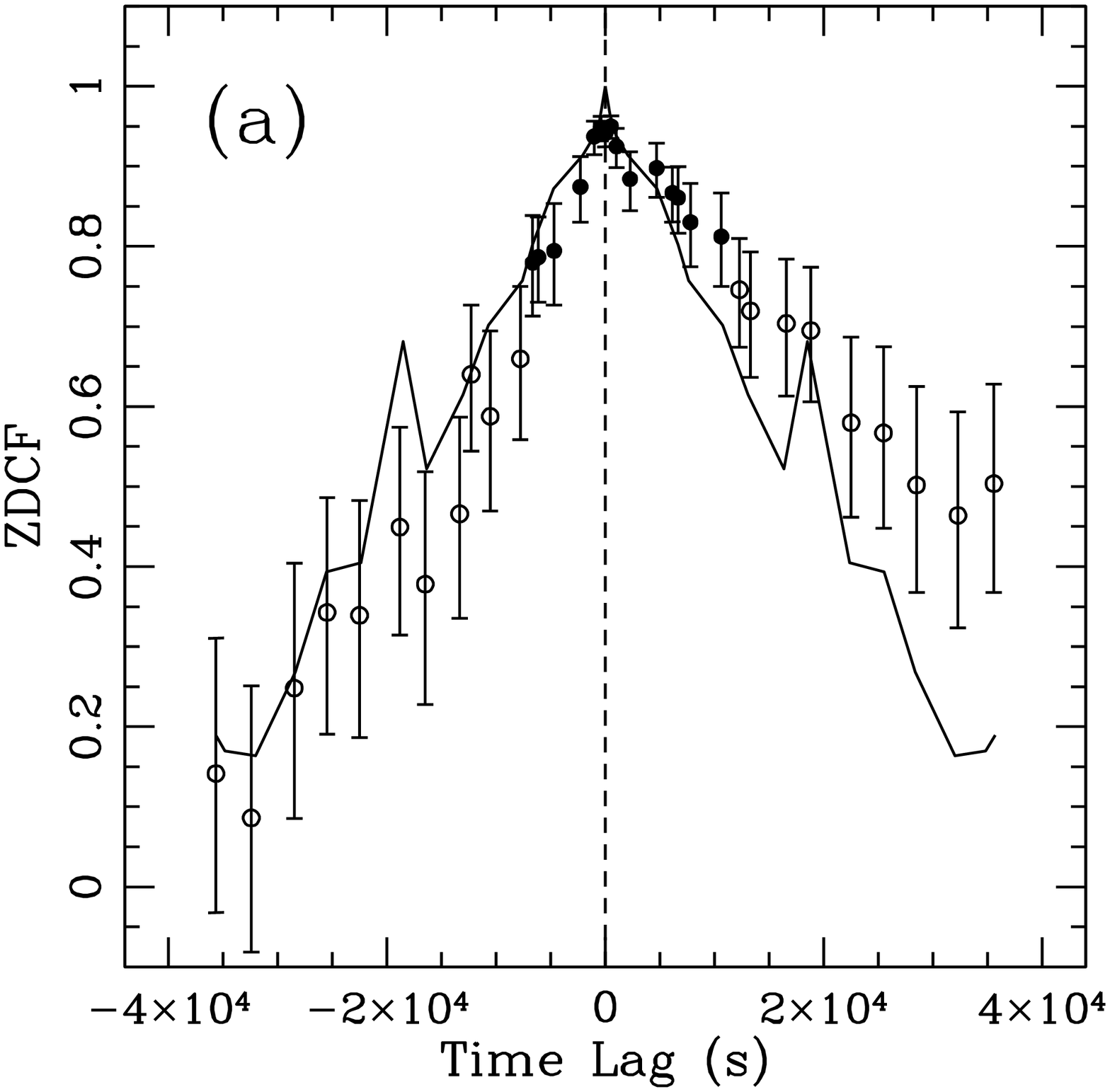,width=8cm,height=8cm} \hspace{8mm}
\psfig{file=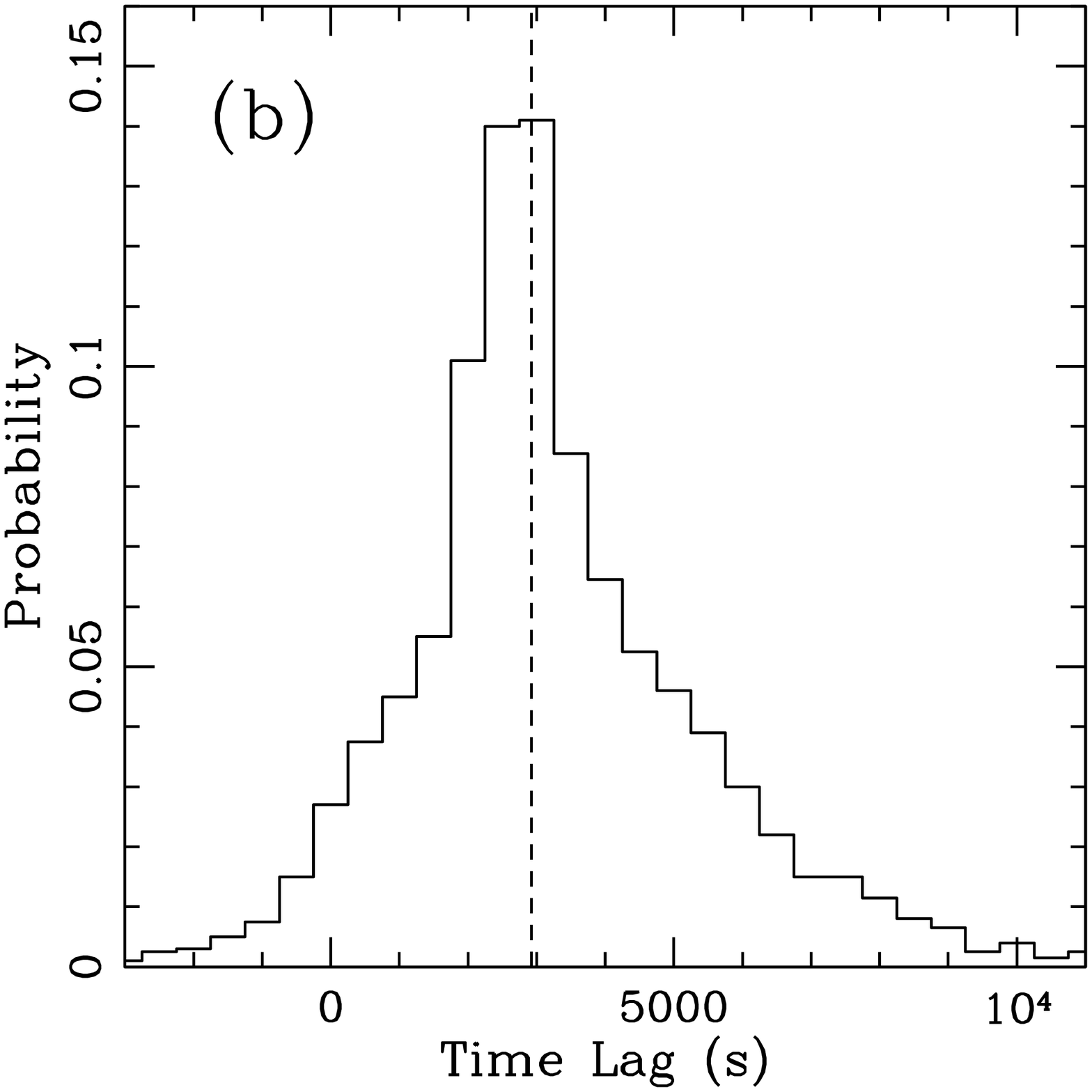,width=8cm,height=8cm} }
\end{tabular}
\caption{ (a) ZDCF (the LECS 0.1--2~keV light curve versus the MECS
2--10~keV one) clearly shows asymmetry towards positive lags while it peaks 
around zero lag. Such asymmetry is easily visible when one compares the 
ZDCF with the 2--10~keV ACF (solid line). The solid circles are the ones 
used to calculate the centroid of the ZDCF, which yield a hard lag of 
$\tau_{\rm cent} = 2.87$~ks. 
(b) The corresponding CCPD of (a) built from the FR/RSS simulations, 
of which the median with $68\%$ confidence level is 
$\tau_{\rm cent} = 2.92^{+2.43}_{-1.50}$~ks. 
The vertical dashed line indicates the median of the CCPD. 
}
\label{fig:zdcf}
\end{figure*}
%-------------------------------------------

The ZDCF of the 0.1--2 versus 2--10~keV bands is shown in 
Figure~\ref{fig:zdcf}a. The light curves are binned at 512~s resolution. 
A positive lag indicates that the higher energy emission lags the 
lower energy one. One can see that the ZDCF is clearly asymmetric towards 
the positive lags, while the peak of the ZDCF is near the zero lag. For 
illustration, we also show in Figure~\ref{fig:zdcf}a (solid line) 
the auto-correlation function (ACF) of the 2--10~keV light curve. 

It is obvious that there are ambiguities associated with determining such 
CCF result, in particular for such small time lag. Three techniques to  
interpret a CCF result exist in the literature: (1) using the time lag 
corresponding to the actual maximum value ($r_{\rm max}$) of the CCF, 
$\tau_{\rm peak}$; (2) computing the centroid of the CCF over time lags 
bracketing $r_{\rm max}$, $\tau_{\rm cent}$; (3) fitting the CCF with a 
function (e.g., Gaussian) to find the location of the peak, 
$\tau_{\rm fit}$, which is the technique used in Fossati \etal (2000a). 
However, $\tau_{\rm peak}$ is not suitable for very small lag compared
to the duration of the time series, because it suffers from statistical
uncertainties, depends on the binning patterns of both the light curves and
the CCF, and does not consider asymmetries of the CCF that is usually seen
in AGN variability. Moreover, because of complex CCF shape in this case, 
$\tau_{\rm fit}$ is not applicable either. We therefore only quote 
$\tau_{\rm cent}$ throughout the paper. 

In Figure~\ref{fig:zdcf}a, the solid circles indicate all points with 
$r$ in excess of $0.8r_{\rm max}$, which are used to calculate the centroid 
of the ZDCF (Peterson \etal 1998). We obtain $\tau_{\rm cent} = 2.87$~ks.
We then perform FR/RSS simulations to evaluate its significance. As described 
in Peterson \etal (1998), a simulation is deemed to have succeeded if 
$r_{\rm max}$ between the two simulated time series is significant at a 
level of confidence greater than $95\%$. For each succeeded trial, 
$\tau_{\rm cent}$ is recorded. After 2000 runs of successful realizations, 
we determine the median of $\tau_{\rm cent}$ with $68\%$ confidence range 
by integrating the CCPD (Figure~\ref{fig:zdcf}b). The simulations suggest 
$\tau_{\rm cent} = 2.92^{+2.43}_{-1.50}$~ks.

\begin{figure}
\psfig{file=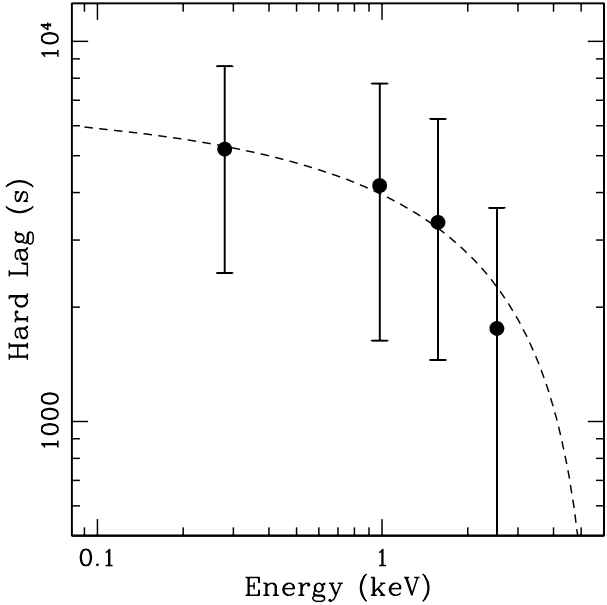,width=8cm,height=8cm}
\caption{ Energy dependence of the hard lag (median of the CCPD) of 
the 3.2--10~keV photons versus the lower energy ones. The two-sided 
error bars indicate $68\%$ confidence range with respect to the median 
of the CCPD. The energies indicate the logarithmically averaged energies 
of the energy bands by taking into account gradually spectral steepening of 
the source. The dashed line indicates the best fit with the energy-dependent 
timescale of a diffusive particle acceleration model (see 
\S\ref{sec:disc:implications} for details). }
\label{fig:lage}
\end{figure}

One obvious issue whether the time lags depend on photon energies can
have some important consequences: such energy-dependence may be indeed
expected by the energy-dependent acceleration/cooling timescale of
relativistic electrons responsible for the observed X-rays. To examine this
possibility, we divide the 0.1--10~keV energy range into 5 bands by taking
into account each band having similar photon statistics, i.e., 
0.1--0.78~keV, 0.78--1.23~keV and 1.23--2~keV (LECS); and 2--3.2~keV,  
3.2--10~keV (MECS). Using the FR/RSS simulations, the time lags of
the first four lower energy bands are estimated with respect to the
3.2--10~keV energy band. The simulations suggest hard lag for each case. 
The results are shown in Figure~\ref{fig:lage} as a function of 
photon energies.
Due to the quality of the data sets used, the errors are quite large, and 
the possibility of a constant hard lag with no energy-dependence cannot be 
excluded. Nevertheless, the hard lags appear to be energy-dependent, 
in the sense that the hard lags of the 3.2--10 keV photons increase 
with the decreasing energies of the comparison softer photons. 

%-------------------------------------------
%-------------------------------------------

		%--------------------%

\section{Fourier Frequency-dependence of Time Lags} \label{sec:flag}

If $X_s(f_i)$ and $X_h(f_i)$ indicate the Fourier transforms of two 
statistically independent but concurrently measured light curves $x_s(t)$ 
(soft energy band) and $x_h(t)$ (hard energy band) with N evenly sampled 
observations ($t=1\Delta t, 2\Delta t, \ldots, N\Delta t$ with $\Delta t$ 
being the bin size of the samples; and $f_i = i/(N\Delta t), 
i=1,2,\ldots,N/2$, is the Fourier frequencies), the cross-spectrum between 
$x_s(t)$ and $x_h(t)$ is defined as (e.g., Nowak et al. 1999; Papadakis
et. al. 2001)
\beq
I_{s,h}(f_i) = X^*_s(f_i) X_h(f_i) \,, 
\label{eq:crossspectrum}
\eeq
where $X^*_s(f_i)$ is the complex conjugate of $X_s(f_i)$. If $X_h(f_i)$ 
(or $X_s(f_i)$) is substituted by $X_s(f_i)$ (or $X_h(f_i)$), 
equation~(\ref{eq:crossspectrum}) becomes the auto-spectrum, i.e., the 
spectral power density (PSD) of $x_s(t)$ (or $x_h(t)$). Unlike PSD, the 
cross-spectrum is a complex function, its argument is defined as the 
phase spectrum between $x_s(t)$ and $x_h(t)$,   
\beq
\phi_{s,h}(f_i) = { \rm tan^{-1} }
        \left [ \frac{ {\rm Im} \{ I_{s,h}(f_i) \} }
         { {\rm Re} \{ I_{s,h}(f_i) \} } \right ] \,,
\label{eq:phase}
\eeq
where ${\rm Re}\{I_{s,h}(f_i )\}$ and ${\rm Im}\{I_{s,h}(f_i)\}$          
indicate the real and imaginary part of $I_{s,h}(f_i)$, respectively. 
$\phi_{s,h}(f_i)$ indicates the phase shift, 
$\phi_s(f_i)-\phi_h(f_i)$, between the fluctuations 
in $x_s(t)$ and $x_h(t)$ at frequency $f_i$.
The corresponding time shift (lag) is given by
\beq
\tau_{s,h}(f_i) = \frac{\phi_{s,h}(f_i)}{2\pi f_i} \,,
\label{eq:lag}
\eeq
which gives Fourier frequency-dependent time lags, i.e., the lag spectrum, 
between the variations of the two time series. 

		%---------------%

\subsection{The observed lag spectrum} \label{sec:flag:results}

%------------------------------------
\begin{figure}
\psfig{file=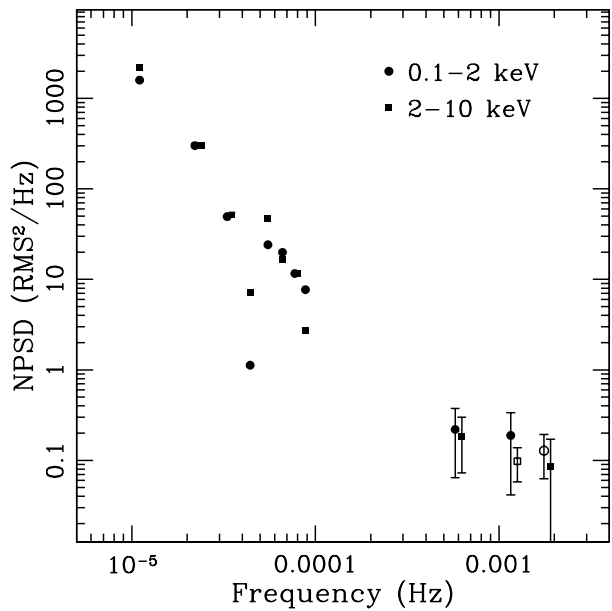,width=8cm,height=8cm}
\caption{ Normalized power spectral density (NPSD) of the 0.1--2~keV and
the 2--10~keV light curves. The noise power has been subtracted. 
The two open symbols indicate negative powers. For clarity, small 
frequency-offsets have been applied. There are no error estimates 
for the powers (derived from orbital period-binned light curves) 
at $f < 10^{-4}$~Hz, while the powers with errors at 
$f > 10^{-4}$~Hz are derived from the 12 continuous segments of the 
256~s binned light curves. 
}
\label{fig:npsd}
\end{figure}
%----------------------------------------

We extract two pairs of light curves, the 0.1--2 (LECS) versus
2--10~keV (MECS), and the 2--3.2 versus 3.2--10~keV (MECS),
respectively. To ensure that the cross-spectrum method is appropriate, 
evenly sampled data sets are 
required. We thus use for the analysis two bin sizes corresponding 
to the \sax\ orbital period and 256~s, respectively. The main goal 
of such analysis is to examine the results evaluated with the 
CCF methods (see \S\ref{sec:lage}), and to investigate 
whether or not the source shows frequency-dependent time lags. 

There are $N=16$ points in the orbit-binned light curves. We first 
estimate individual PSD and cross-spectrum of the two light curves 
of a pair. The phase spectrum (equation~\ref{eq:phase}) is  
calculated on the basis of the real and imaginary part of the 
cross-spectrum, and then the lag spectrum (equation~\ref{eq:lag}). 
This procedure gives PSDs and lag spectrum of a pair at 8 
frequencies in the low frequency range.

For each continuous data segment of the 256~s binned light curves that 
have no gaps, the PSDs and cross-spectrum of a pair are calculated 
with the same procedure as above. There are 12 such segments (orbits).  
We then combine all of them, sort them in order of increasing frequency 
and group them to 3 (frequency) bins with equal number for both the PSDs  
and the cross-spectrum. The phase spectrum of the pair are then estimated 
with the average cross-spectrum. This procedure yields PSDs and lag spectrum 
of 3 points at high frequency range.

%------------------------------------------------------------------
\begin{figure*}
\begin{tabular}{cc}
\centerline{
\psfig{file=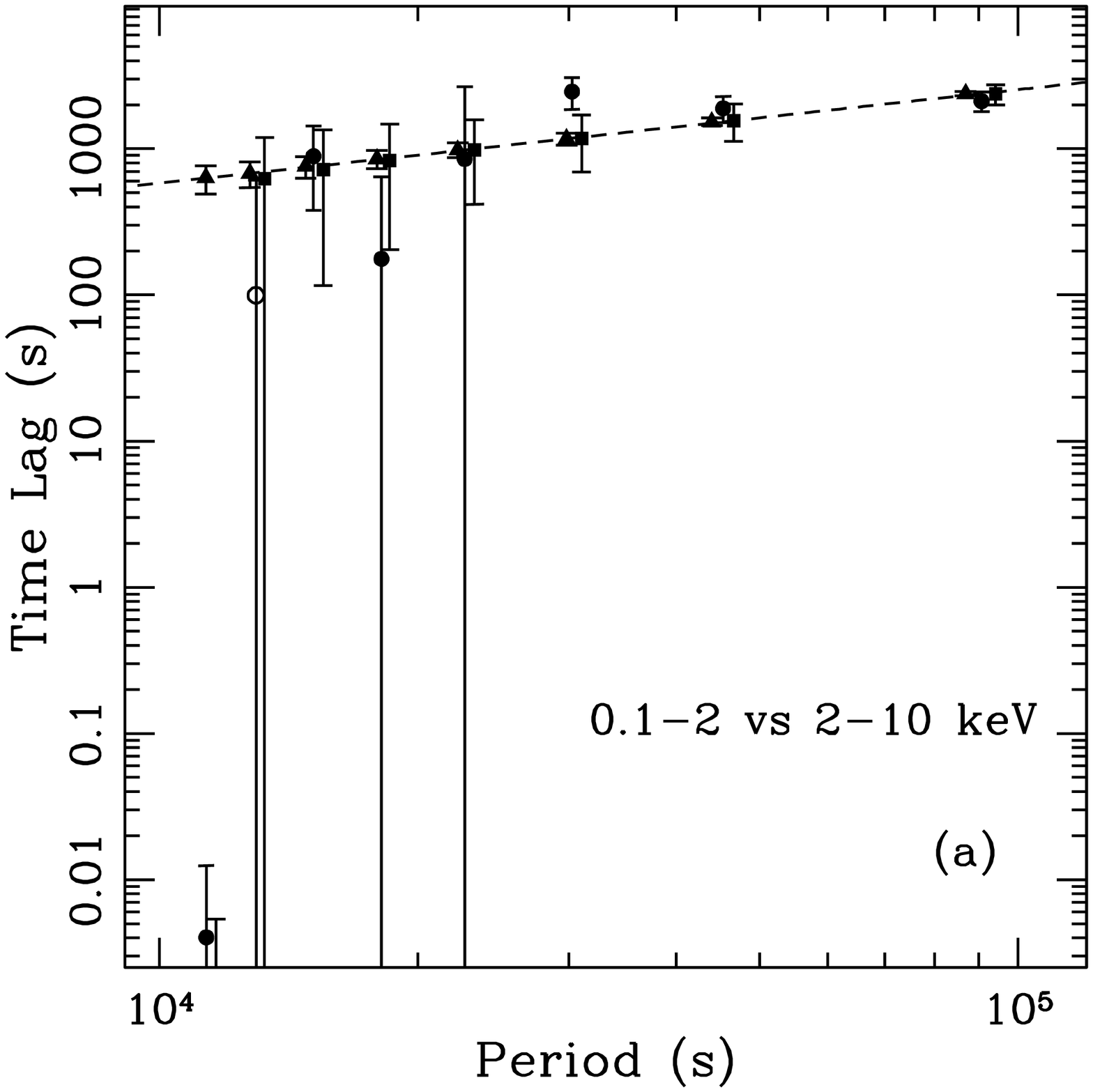,width=8cm,height=8cm} \hspace{8mm}
\psfig{file=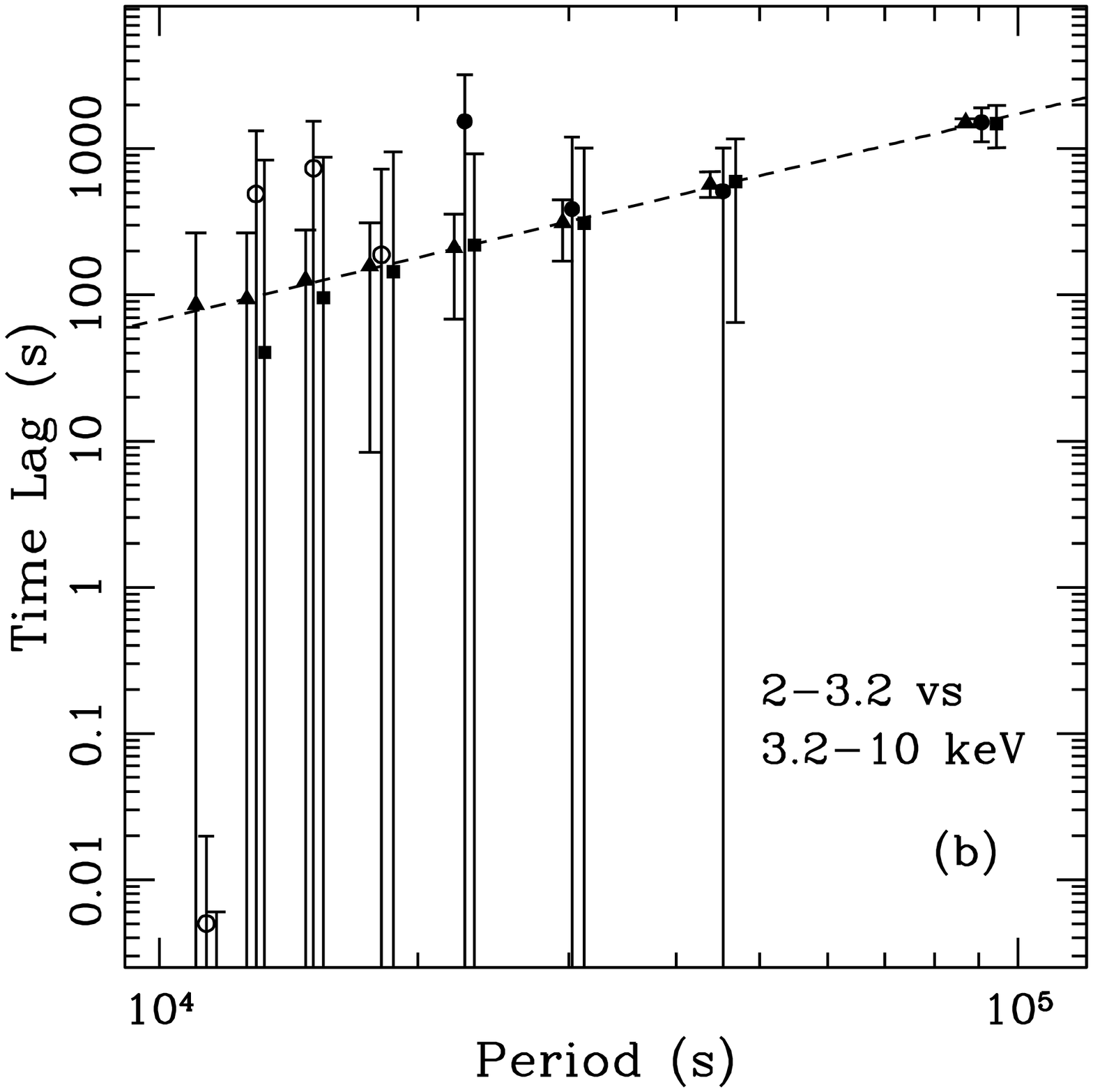,width=8cm,height=8cm} }
\end{tabular}
\caption{ Time lags as a function of Fourier period ($1/f$) derived from  
the cross-spectrum of (a) 0.1--2~keV versus 2--10~keV; (b) 2--3.2~keV 
versus 3.2--10~keV. The light curves are 5670~s-binned. 
The solid (positive) and open (negative) circles are the observed lags 
of which the errors indicate $68\%$ confidence level due to Poisson 
noise with respect to the observed lags. 
The dashed line shows the unweighted best-fit power law model to the 
observed positive (hard) lags by excluding the negative ones. Moreover, 
in (a) the point at shortest period is also excluded, and in (b) only the 
points at the 3 longest periods are used for the fits. The fits do not 
indicate the accurate phase spectrum of the source, which is only used 
for the purpose of simulations. 
Solid squares (5670~s binned light curves) and triangles (256~s binned 
light curves) are the simulation lags (the medians with $68\%$ 
confidence levels) by applying the assumed phase 
spectrum (dashed lines) to the second light curves of the faked pairs 
(See \ref{sec:flag:sim:phase} for details). For clarity, only 8 points 
at the longest periods are shown for the case of the 256~s-binned light 
curves, and period offsets have been applied, and because of very small 
value the point at the shortest period from 5670~s binned light curves is 
only seen with part of the upper error bar. 
The simulations show that densely sampled light curves can well recover 
the assumed phase spectrum that can be distorted by sparsely sampled 
light curves.  }
\label{fig:flag}
\end{figure*}
%--------------------------------------------------------------------

In practice, it is customary to divide the light curves into many
segments, compute the PSDs and cross-spectrum for each of them,
average PSDs, ${\rm Re}\{I_{s,h}(f_i)\}$ and
${\rm Im}\{I_{s,h}(f_i)\}$, and/or rebin them by averaging say $m$ 
consecutive frequency bins to obtain an estimate of the PSDs and 
phase spectrum with statistical uncertainty. 
However, our \lcs\ can not match these conditions because of the 
short duration (i.e., only one single flare was sampled with 16 data 
points for the orbital period-binned light curve). We believe 
that this choice is compulsory for the analysis of the TeV sources 
because the variability properties as discussed in the introduction 
differ from flare to flare. Accordingly, the
statistical uncertainties of the results can not be
accounted for by dividing the \lcs\ into segments and rebinning them by
averaging consecutive frequency bins. Therefore, at least for the
orbital period-binned \lcs , the uncertainties on the PSDs and time lags 
can not be evaluated according to equation [16] of 
Nowak \etal (1999). It means that there are no error estimates for the 
results obtained with the orbital period-binned light curves. 
Even though the results obtained with the 256~s binned light curves 
have error estimates, low S/N ratio limits the statistical significance.
     
The resulting PSDs in the 0.1--2 and 2--10~keV energy band are shown 
in Figure~\ref{fig:npsd} (normalized and noise power subtracted as 
in Zhang \etal 2002). One can see that the two normalized PSDs (NPSDs) 
follow power-law shapes with very steep slope, in the sense that the 
power quickly decreases with increasing frequency, a strong 
\textit{red noise} feature. There is no clear difference between 
the two NPSDs.

Due to the limited statistics of the 256~s binned light curves, we only show 
the observed lag spectrum derived from the orbital period-binned 
light curves. As commented above, we are discussing here the error-free 
results. The issues on the errors will be explored with detailed simulations 
in the next section (\S\ref{sec:flag:simulations}). Figure~\ref{fig:flag} 
shows the lag spectrum as a function of Fourier period ($1/f)$. 
A positive value indicates that the component at frequency $f$ in the higher 
energy band is delayed with respect to the same component in the lower 
energy band.
The 0.1--2/2--10~keV pair (Figure~\ref{fig:flag}a) shows 7 positive 
(solid circle) and 1 negative (open circle) lags, while the 
2--3.2/3.2--10~keV pair (Figure~\ref{fig:flag}b) shows 4 positive and 
4 negative values.
The dependence of the positive (hard) lags on the periods seems not to be 
unique: the amplitudes of the lags seem to remain constant or increase  
with periods. For the four longest periods, all lags are
positive with values of $\sim 1000$--$2500$~s and $\sim 500$--$1500$~s 
for the 0.1--2/2--10~keV and the 2--3.2/3.2--10~keV pairs, respectively. 
For the other four periods, the situation is more complicated for 
both cases. We also mention that the time lags from the 256s-binned 
pairs are relatively small with either positive or negative signs. 

\subsection{Simulations} \label{sec:flag:simulations}

It is obvious that short duration, periodic gaps, sparsely sampling, 
low S/N ratio and red noise character of the data sets deserve 
simulations in order to determine the significance and the uncertainties 
of the error-free results obtained in last section 
(\S\ref{sec:flag:results}). 

\subsubsection{Poisson noise}\label{sec:flag:sim:poisson}

\begin{figure} 
\centerline{
\psfig{file=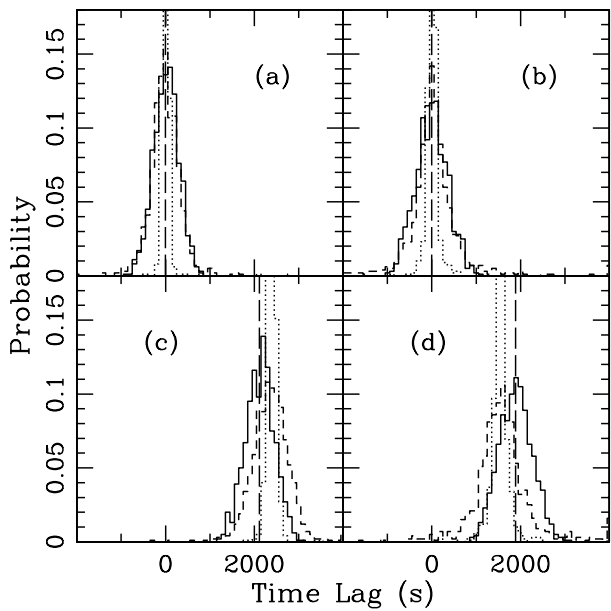,width=8cm,height=8cm}}
\caption{
Probability distributions of the simulation lags at the 
two lowest frequencies (i.e., $f=1.1\times 10^{-5}$ Hz (a,c); 
$f=2.2\times 10^{-5}$ Hz (b,d) ).
(a,b): 
the solid lines represent the effects of Poisson noise (using the observed 
5670~s binned 2--10~keV light curve as a template);  
the short dashed/dotted lines indicate the effects of red noise (the 
PSD is assumed to be a power-law model of slope 2.5). The short dashed 
lines indicate that the faked pairs have the same sampling pattern, mean, 
variance and Poisson noise as the observed 5670~s binned 0.1--2/2--10~keV 
pair. 
The dotted lines are the same as the short dashed ones except for sampling  
the faked pairs with 354 points of 256~s resolution, and applying 
the average errors of the observed pair; 
The vertical long dashed lines indicate zero lags by construction. 
%%%
(c,d): 
the solid lines are the effects of Poisson noise on the observed 5670~s 
binned 0.1--2/2--10~keV pair. The vertical long dashed lines indicate 
the observed lags; 
the short dashed/dotted lines correspond to those of (a,b), respectively, 
except for the phase spectrum of the second light curve of the faked pair 
being delayed on the basis of the best-fit power-law model to the observed 
0.1--2/2--10~keV error-free phase spectrum as shown in 
Figure~\ref{fig:flag}a. 
The offsets between the peaks of the solid (also the vertical dashed 
lines) and the short dashed/dotted lines indicate the deviations between 
the observed and the assumed phase spectrum at the the same frequencies.
Note that in all cases the peaks of the short dashed and the 
dotted lines overlap each other but the latter has been out 
of the plotting range.
}
\label{fig:lagprob}
\end{figure}

We first perform simulations to investigate the effects of Poisson
noise (relating to S/N ratio) to the measured lag  
at each frequency. We use the observed 2--10~keV orbital period-binned 
light curve as a template. We create a pair of light curves by Gaussian 
randomly redistributing the fluxes of the light curve using the FR method 
as adopted in \S\ref{sec:lage} on the basis of the quoted errors on the 
real light curve. By construction, the pair is identical except for the 
effects of Poisson noise, zero lag is expected at each frequency except 
for a random component contributed by Poisson noise. 
The pair is analysed with the cross-spectral method in the same 
way as a real pair, and the lags at each frequency are 
recorded. After 1000 runs of this process with different sets of random 
number, the probability distributions (similar to CCPD) are built for the 
lag at each frequency. Figures~\ref{fig:lagprob}a,b (solid lines) show such 
distributions at the two lowest frequencies (i.e., longest periods). As 
expected, one can see that the lag probability distributions peak 
at zero. Moreover, the medians of the simulation lags are very close to 
zero at any frequency, and they are not dominated by positive values as 
observed in Figure~\ref{fig:flag}a. 

As a result, Poisson noise does not result in systematic bias to the
observed lags while it introduces an uncertainty. We then
investigate how large uncertainty it can introduce. This has been done in
the same way as above, but we use light curves at different energy bands
rather than those at same energy band. We repeat the above process with 
the pairs of the orbital period-binned 0.1--2/2--10~keV and 
2--3.2/3.2--10~keV light curves, respectively. Figure~\ref{fig:lagprob}c,d 
(solid lines) show the probability distributions of the lags 
at the two lowest frequencies of the 0.1--2/2--10~keV pair, respectively. 
One can see that the peaks of the lag probability 
distributions overlaps with the observed ones (vertical dashed lines) 
while Poisson noise does introduce an uncertainty to the 
observed lags. Given the sampling 
pattern (see \S\ref{sec:flag:sim:rednoise}), the degree of such  
uncertainty depends on the S/N ratios of the two light curves. 
After comparing the probability distributions shown with solid lines in
Figure~\ref{fig:lagprob}c,d with those shown with solid lines in
Figure~\ref{fig:lagprob}a,b, we can conclude 
that the observed lags are not an artifact of Poisson noise. 
In Figure~\ref{fig:flag}, the error bars on the solid/open circles (i.e., 
the observed values) show the $68\%$ confidence level of uncertainties 
introduced by Poisson noise with respect to the observed lags. It can been 
seen that the relative uncertainty generally increases with decreasing period. 

However, it is important to point out, as we will show below, that most 
of such uncertainties due to Poisson noise as shown above are mainly 
related to sparsely sampling character of the light curves when one 
compares the simulation results obtained with faked light curves with 
different resolutions (given that the light curves have same length and 
S/N ratio). The probability distributions can be significantly broadened 
for sparsely sampled light curves, which underestimates the significance 
of the observed lags.  

%---------------------------------------------------------------------

\subsubsection{Red noise}\label{sec:flag:sim:rednoise}

Figure~\ref{fig:npsd} has shown that the source shows strong
red noise variations. The unweighted fits with a power-law model to the 
error-free PSDs suggest a steep slope of about 2.5 for both the 
0.1--2~keV and the 2--10~keV PSDs. It is therefore important to 
investigate the effects of such strong red noise character on the 
detected lags. To do so, we use randomly generated light curves rather 
than the observed ones.

Inverse Fourier transformation from frequency domain to time domain is 
usually performed to recover the light curves from an assumed PSD 
model. We use the algorithm of Timmer \& K\"{o}nig (1995), which randomizes 
both the amplitude and the phase of the PSD at each Fourier frequency, and 
is superior to the commonly used ``phase only randomisation'' approach. 
Throughout this work, we assume a power-law PSD model with a slope 2.5 
appropriate for the source. We consider three cases as follows.

Case 1:  with one set of Gaussian distributed random number, we first 
fake a light curve resembling the sampling pattern of the real orbital
period-binned light curves, i.e., the faked light curve has 16 points
with 5670~s resolution. We then make this faked light curve become to a
pair by scaling it to have the same mean and variance as the real
0.1--2~keV and 2--10~keV light curves, respectively. The two light curves 
of the faked pair is
then Gaussian randomly redistributed using the FR technique on the basis
of the quoted errors on the real 0.1--2~keV and 2--10~keV light
curves, respectively. The two light curves of the faked pair are therefore 
identical except for differing mean, variance and Poisson noise, zero lag 
is then expected at each frequency. The cross-spectral analysis is
performed on the faked pair as a real pair, from which the lags 
are recorded for each frequency. Using different sets of random number, 
we repeat this process by a number of 1000 to build the probability 
distributions of simulation lags at each frequency. We show 
in Figures~\ref{fig:lagprob}a,b (short dashed lines) such distributions at 
the two lowest frequencies. One can see that the lag probability 
distributions due to red noise are similar to those due to Poisson 
noise only. It means that given the same sampling pattern, the lag 
uncertainty is mainly associated with Poisson noise rather 
than red noise. Moreover, same as the effects of Poisson noise, 
the medians of the simulation lags are likely very close to zero 
at any frequency, and they are not positive-values dominated as plotted 
in Figure~\ref{fig:flag}a either. We conclude that the detection of 
lags from the real pairs of light curves cannot be the result 
of red noise character of the flare.

Case 2: we repeat the above process with faked light curves having 354 
points with 256~s resolution, producing total duration of light curves 
close to the ones having 16 points with 5670~s resolution. We adopt the 
same scaling factors as Case 1, and the average Possion noise of the 
real orbital period-binned 0.1--2 and 2--10~keV light curves, which 
is similar to those of Case 1, is applied when redistributing the fluxes 
of the faked light curves, respectively. This set of simulations is used 
to investigate the effects of different sampling patterns for the 
deterministic duration of light curves. The results are shown in 
Figures~\ref{fig:lagprob}a,b as the dotted 
lines. One can clearly see that densely sampled light curves yield 
much narrower lag distributions than the sparely sampled ones do. It means, 
given the same Poisson noise level, that more sparsely sampled light 
curves introduce larger uncertainty. 

Case 3: Case 2 is repeated with the exact same sets of faked pairs 
except for doubling the average Poisson noise level. The results show,  
given sampling pattern and duration, that the larger is Poisson noise 
level, the larger is the uncertainty.

\subsubsection{Phase-shifted red noise}\label{sec:flag:sim:phase}

Having established that the observed (phase) lags are not spurious results of 
Poisson and red noise on the observed data sets, we then investigate what 
are the ``errors'' associated with the observed lags. Because the real data 
sets do not allow us to get proper estimates of the errors directly, we still 
use faked light curves to investigate how large the errors can be introduced 
by Poisson and red noise together.

We repeat the three cases in last section (\S\ref{sec:flag:sim:rednoise}) 
in the same way, i.e., the same sets of random number, the same 
scaling factors and Poisson noise levels. The only difference is that 
one set of random number directly generate a pair of light curves 
rather than one light curve. For each faked pair, the phase spectrum 
of the second light curve is delayed with respect to the first one on 
the basis of the observed phase spectrum. To do so, an unweighted fit with 
a power-law model is performed to the observed error-free lag spectrum 
of the 0.1--2/2--10~keV pair (Figure~\ref{fig:flag}a). The best fit 
suggests $\tau(t) = 1.60 t^{0.64}$ after excluding the negative value and 
the one at the shortest period, which is shown in Figure~\ref{fig:flag}a  
(the dashed line). However, we emphasize that such a fit does not really 
represent the accurate phase spectrum of the flare, it is just used as a 
reasonable model to investigate the uncertainty and significance of the 
observed results. 

The probability distributions of the simulation lags at the two lowest 
frequencies are shown in Figures~\ref{fig:lagprob}c,d, which correspond to 
those of Figures~\ref{fig:lagprob}a,b  (i.e., the Case 1 and 2 of 
\S\ref{sec:flag:sim:rednoise}), respectively. The short dashed and dotted 
lines show the results derived from the 16 (5670~s resolution) 
and 354 (256~s resolution) points sampled light curves, respectively. One 
can see that the assumed lags are easily recovered. The peaks of the 
lag distributions for the two cases overlap each 
other while the denser sampled light curves yield narrower distributions. 
Note that the small offsets between the peaks of the simulation 
lags (short dashed/dotted lines) and those of the observed ones (solid 
lines; also see the vertical dashed lines) indicate the deviations of the 
latter from the best-fit phase spectrum. 

In Figure~\ref{fig:flag}a, we show the medians of the simulation lag 
distributions with $68\%$ confidence level at all the 
8 frequencies. The solid squares and triangles represent the results from 
the case of 16 and 354 points sampled light curves, respectively. For 
clarity, the results from the 354 points sampled light curves are shown 
at only the 8 lowest frequencies that correspond to those of the 16 points 
sampled ones. It can be seen that the simulations of the 16 points 
sampled light curves recover well the observed lag spectrum 
at the longest periods. The simulation lags are 
very similar to the observed ones. Moreover, the dispersions on the 
simulation values are also similar to those on the observed ones due to 
Poisson noise only, especially for the two lowest frequencies. This 
suggests that the uncertainty due to Poisson noise may account for the 
uncertainty of the observed lags, which resembles the
uncertainty on the CCF lags with the FR/RSS simulations. 
By comparing the simulation results obtained with 16- and 354-points 
light curves, one can see that sparsely sampled light curves give rise to 
lag spectrum with much larger uncertainties and distort the true lag 
in the shortest period.     

We repeat the above processes for the observed 2--3.2/3.2--10~keV pair of 
light curves. We also fit the observed error-free (i.e., unweighted)  
lag spectrum (Figure~\ref{fig:flag}b) with a power-law model. 
The fit is performed using only points at the 3 lowest frequencies.
The best fit suggests $\tau(t) = 1.62\times 10^{-4}t^{1.41}$,  which is 
shown in Figure~\ref{fig:flag}b (dashed line).  
The simulation results are shown in Figures~\ref{fig:flag}b 
with solid squares and triangles corresponding to the faked light curves 
having 16 and 354 points, respectively. 
Similar to the 0.1--2/2--10~keV case, one can clearly see 
how the sparsely sampled light curves distort the assumed phase spectrum 
in the sense of both the trends and uncertainties.
We conclude that the observed dependence of the lag on 
frequency has been distorted by the sparsely sampled data sets.  

The results corresponding to the Case 3 of \S\ref{sec:flag:sim:rednoise} 
(i.e., using larger Poisson noise level) is that the assumed  
phase spectra are maintained but with larger 
dispersions, demonstrating the main effects of low S/N ratios.

As a summary, our simulations show that the observed phase spectra are 
intrinsic to the source but distorted by the 
available data quality. The future continuous observations with high 
resolution and S/N ratio will be able to determine whether or not 
the time lags in TeV blazars are Fourier frequency-dependent. 

		%--------------------%

\section{Discussion}\label{sec:disc}

\subsection{CCF lag versus phase lag} \label{sec:disc:summary}

We reexamined the discovery of hard lags in the X-ray emission during 
a large flare of \mkn\ observed with \sax . It was investigated 
in several ways. The presence of hard lag character is firstly shown
with the anticlockwise track in the diagram of hardness ratio versus
count rate. We then quantified the hard lags with the CCF 
method. Finally, we performed an cross-spectral analysis with detailed 
simulations to evaluate the effects of Poisson and red noise.
 
The CCF is quite complicated and clearly shows asymmetry and no peak at a  
lag different from zero. The centroid method suggests that 
the 2--10~keV emission lags the 0.1--2~keV one by about 3000~s, and the 
hard lags may be energy-dependent. However, given the data quality in hand 
and ambiguities in interpreting the complicated shape of the CCF, these 
results are associated with quite large errors obtained with the FR/RSS 
simulations. Moreover, the condition that only the CCF points with $r$ 
in excess of $0.8r_{\rm max}$ should be used in the determination of 
$\tau_{\rm cent}$ is just an empirical rule, had one chosen to consider only 
the few points around the maximum of the CCF (which is close to zero lag ) 
in order to compute $\tau_{\rm cent}$, then the observed ``lag''  
would be close to zero. So, it could be that the ``important/real'' fact is 
that the CCF indeed has a maximum at $\sim$ zero lag, and the estimated 
lag in \S\ref{sec:lage} is just a result of the CCF asymmetry. In other 
words, the observed asymmetric CCF is difficult to be interpreted   
since the lag determined from such CCF is not uniquely determined (i.e., it 
is not easy to understand what exactly this``lag'' means). Therefore, in this 
case the CCF may not address whether there is a ``lag'' or not, while 
certainly asymmetric CCF presents significant evidence that the hard photons 
are delayed with respect to the soft ones. 

One possibility that may produce the complicated CCF shape is that 
the peak of the CCF is possibly dominated by the fastest variations while 
the asymmetry may be caused by the relative delays of the variations on the 
longer timescales (e.g., Papadakis et al. 2001). This fact implies that the 
variations of different periods in various energy bands may not be delayed 
by the same amount. This further implies that the CCF may always 
not be the applicable method to correctly address such a situation.
The observed CCF lag in such a situation may be difficult to 
understand/interpret, but could be still important.
As an alternative technique to examines the time lag, only the  
cross-spectral technique may actually reveal the ``real'' situation. 
However, the periodic gaps prevent us from performing a full Fourier 
transformation. The orbital period-binned data sets (only 16 points) 
therefore yield the observed lag spectra with rather large uncertainties, 
and the dependence of the lags on frequency is not yet clear. Nevertheless, 
the detection of statistically significant hard lags between the light 
curves at the lowest frequencies, which cannot be the result of red noise 
or Poisson noise effects, assures that the character of hard lag is not 
problematic.
The detailed simulations show that the future higher quality 
data with higher resolution and S/N ratio, which can be available with 
Chandra and XMM-Newton, will be able to construct reliable 
frequency-dependent time lags.

By taking into account the points raised above 
for the CCF, a direct comparison between the lags obtained with the two 
different methods is not applicable in this case. Had the CCF and phase lags 
been uniquely defined, the former should be the average of the latter over 
the probed frequency range. However, whether two light curves with 
frequency-dependent time lags will also show a CCF with a ``fixed lag" is 
not clear yet, in fact it is not clear what this ``CCF lag'' really means. 
The CCF may have a ``fixed'' lag if the lag spectrum were  
frequency-independent, i.e., $\tau(f)=\tau_{\rm CCF}=$ constant.   
The cross-spectral method may therefore be really superior to the commonly 
used CCF methods when one investigates time lags.  

\subsection{Implications of the hard lag} \label{sec:disc:implications}

Intensive long looks of the three TeV blazars with \sax\ and ASCA have 
revealed that the inter-band X-ray time lags differ significantly from 
flare to flare. 
Both soft and hard lags were found, and show dependence on photon
energies. Such complex changes of the time lags in these sources have
been ascribed to relative changes of acceleration and \sy\
cooling timescales, $t_{\rm acc}$ and $t_{\rm cool}$, of relativistic
electrons (e.g., Kirk, Rieger \& Mastichiadis 1998). \tacc\ and
\tcool\ is electron (in turn photon) energy-dependent: it is such 
dependence that might produce the complexities of the observed 
energy-dependent time lags. It is possible that relativistic electrons are
accelerated at shock fronts taking place in jets. The widely discussed
diffusive shock acceleration in particular (e.g., Drury 1983; Blandford \&
Eichler 1987) could be the mechanism responsible for electron acceleration
during the flare of Mrk~421 studied in this work. In such hypothesis, 
\tacc\ and \tcool\ in the observer's frame are expressed as 
(Zhang \etal 2002)  
\beq
t_{\rm acc}(E) = 9.65 \times 10^{-2} (1+z)^{3/2} \xi 
		B^{-3/2} \delta^{-3/2} E^{1/2} \quad {\rm s} \,,
\label{eq:tacckev}
\eeq
\begin{equation} 
t_{\rm cool}(E) = 3.04 \times 10^{3} (1+z)^{1/2} B^{-3/2} \delta^{-1/2}
	E^{-1/2}  \quad {\rm s} \,,
\label{eq:tcoolsykev}
\end{equation}
where $E$ is the observed photon energy in unit of keV, 
$B$ and $\delta$ are the magnetic field and the Doppler factor of the
emitting region, respectively. $\xi$ is the acceleration parameter
indicating the acceleration rate of electrons. One can see that both
$t_{\rm acc}$ and $t_{\rm cool}$ depend on photon energies but
in the \textit{opposite} sense, i.e., the lower energy photons
correspond to electrons with shorter acceleration but longer cooling
\tss\ than the higher energy photons do. If the acceleration is
instantaneous, (i.e., $t_{\rm acc} \ll t_{\rm cool}$, which corresponds
to the case of instantaneous injection of electrons), cooling dominates
the variability and the soft lag is expected as 
\begin{equation}
\tau_{\rm soft}(E) = t_{\rm cool}(E) - t_{\rm cool}(E_{0})
\label{eq:soft}
\end{equation}
If instead the acceleration is slower and comparable with cooling 
(i.e., $t_{\rm acc} \sim t_{\rm cool}$), it takes longer time
for higher energy electrons to accelerate to the radiative
energy, acceleration dominates the system and the  
hard lag is expected as 
\begin{equation}
\tau_{\rm hard}(E) = t_{\rm acc}(E_{0}) - t_{\rm acc}(E)
\label{eq:hard}
\end{equation}
where $E_{0}$ is the referenced energy at which the lag is
measured, and $E<E_{0}$. One can see that the observed inter-band hard
(soft) lag is defined as the difference of $t_{\rm acc}$ ($t_{\rm
cool}$) at different energies, from which the physical parameters of
the emitting region can be constrained (see Zhang \etal 2002 for details). 

Moreover, it is interesting to note that \tacc\ and \tcool\ have the same
dependence on $B$, the ratio of \tacc\ to \tcool\ is thus independent
of $B$. We then have 
\begin{eqnarray}
    t_{\rm acc}(E)/t_{\rm cool}(E) 
  = 3.17 \times 10^{-5} (1+z) \xi \delta^{-1} E \nonumber \\
  = 0.32 \times (1+z) \xi_5 \delta_1^{-1} E \,,
\label{eq:ratio}
\end{eqnarray}
where $\xi_5$ and $\delta_1$ are in unit of $10^5$ and $10$,
respectively. 
One can see that whether a flare shows soft or hard lags at the considered 
energy depends on the (energy-dependent) ratio of 
$t_{\rm acc}/t_{\rm cool}$ (see also Zhang \etal 2002 for an extensive
discussion). If one knows $t_{\rm acc}/t_{\rm cool}$ at particular 
energies (e.g., the maximum synchrotron emitting energy, $E_{\rm max}$,
 where 
$t_{\rm acc}/t_{\rm cool} \sim 1$; or the energy where the soft lag changes  
to the hard lag along the studied energy band), 
equations~(\ref{eq:soft})--(\ref{eq:ratio}) can uniquely determine the 
main parameters of the emitting region, i.e., $B$, $\delta$ and $\xi$.  

It is worth noting that the effects related to particle acceleration 
(i.e., the hard lag from point view of observation) are observable only 
if the system is observed at energies close to $E_{\rm max}$ where 
$t_{\rm acc} \sim t_{\rm cool}$ (Kirk \etal 1998), indicating a relatively 
low acceleration rate. It can been seen from Equation~(\ref{eq:ratio}) 
that $\xi$ 
is the key parameter to modulate the variability pattern of a flare
since $\delta$ is thought to be always the order of $\sim$~10--25. The
value of $\xi$ is poorly known, but changes of $\xi$ provide clues on
changes of the shock parameters. Because of the independence of $t_{\rm
acc}/t_{\rm cool}$ on $B$, by assuming $E_{\rm max}$ (i.e., the maximum 
electron energy, $\gamma_{\rm max}$ which can be accelerated by considering 
the balance between acceleration and 
cooling), $\xi$ can be constrained without knowing $B$ (for $\delta$ in
a small range). On the basis of the commonly assumed parameters 
($\gamma_{\rm max} \sim 10^{5}$, $B\sim 0.1$~Gauss, and
$\delta \sim 10$) of the emitting region to account for the SEDs of the 
Tev blazars (e.g, Kirk \etal 1998; Tavecchio \etal 1998; Fossati \etal
2000b; Kino \etal 2002), the characteristic maximum synchrotron energy
as seen in the observer's frame, $E_{\rm max} = 3.7\times 10^6 \delta B
\gamma_{\rm max}^2$~Hz, emitted by the electrons with $\gamma_{\rm max}$, 
is at the order of a few keV. We then
reasonably assume $E_{\rm max} \sim 10$~keV for the flare discussed
here. Using Equation (\ref{eq:ratio}) and assuming that $\delta \sim
10$ and $t_{\rm acc} = t_{\rm cool}$ at $E_{\rm max}=10$~keV, $\xi$ is
estimated to be $ \sim 3.06 \times 10^4$. Moreover, if the 
energy-dependent hard lags shown in Figure~\ref{fig:lage} can be 
interpreted by $t_{\rm acc}(E)$, we can fit it using
Equation~(\ref{eq:hard}). The best fit is shown in Figure~\ref{fig:lage} 
as the dashed line. It can be seen that \tacc($E$) fits reasonably well 
the observed energy-dependent hard lags, supporting the assumption that  
the flare is dominated by \tacc . The best fit gives $B\delta
\xi^{-2/3} = 1.08 \times 10^{-3}$ Gauss. If the above values of
$\delta$ and $\xi$ were assumed, we then have $B \sim
0.11$~Gauss, which is self-consistent with the value used in simulating 
this flare by Fossati \etal (2000b) and the usual range of $B$ obtained by 
fitting the SEDs, i.e., $\sim$ 0.1--0.3 Gauss for the three
TeV sources (e.g., Ghisellini, Celotti \& Costamante 2002)

With the parameters inferred above, in the observer's frame, \tacc\ and 
\tcool\ is $\sim 8.46 \times 10^3$~s at 10~keV; while \tacc\ is 
$\sim$ $2.68 \times 10^3$~s, \tcool\ $\sim$ $2.68 \times 10^4$~s, and 
then $t_{\rm acc} \sim 0.1 t_{\rm cool}$ at 1~keV. One can see from 
Figure~\ref{fig:lc}a that the flare shows quasi-symmetric profile of the  
rise and decay \ts\ of $\sim 5\times 10^4$~s, which may indicates the 
light-crossing time ($t_{\rm cross} \sim R/(\delta c)$) across the 
emitting region. It is then obvious that both \tacc\ and \tcool\ are 
smaller than $t_{\rm cross}$ in the high energy band, while 
$t_{\rm cool}$ becomes comparable to $t_{\rm cross}$ in the soft band. 
Accordingly, the observed emission should be a superposition from 
different parts of the 
emitting region with the electron distribution at different stages of
evolution. Then the time profile of each variation is not directly
observable, but more visible soft lags in the case of \tacc $\ll$
\tcool\ are obtained by taking into account the light-crossing time
effects (Chiaberge \& Ghisellini 1999). By analogy, the light-crossing
time effect introduces observable hard lag in the case of \tacc $\sim$
\tcool .   

\subsection{Comparison with other black hole accreting systems}
\label{sec:disc:comparison}

The X-ray hard lag observed in \mkn\ is phenomenally similar to those 
observed in the X-ray variability of GBHCs and Seyfert galaxies.
The CCFs of GBHCs (e.g., see Maccarone, Coppi, \& Poutanen 2000 for 
Cyg~X--1; Smith \& Liang 1999 for GX~339--4) and Seyfert galaxies (e.g., 
see Lee et al. 2000 for MCG--6-30-15; Papadakis \& Lawrence 1995 for 
NGC~4051; Nandra \& Papadakis 2001 for NGC~7469) also clearly show 
asymmetries towards hard lags. As discussed in \S\ref{sec:disc:summary}, 
this suggests that in these sources the relationships between the variations 
in different energy bands are not simply a ``fixed'' lag either. This has 
led to the fact that recent lag determinations in GBHCs have concentrated 
in Fourier frequency domain. A number of investigations have shown that in 
GBHCs the dependence of hard lag on frequency can approximate as 
$\tau(f) \propto 1/f$ with a break at the lowest frequency range (e.g., 
see Nowak et al. 1999 for Cyg~X--1 ) while such dependence is generally 
far more complicated. The hard lags in GBHCs also depend photon energies  
(Miyamoto et al. 1998). In comparison, the Seyfert galaxy NGC~7469 also 
approximately exhibit $\tau(f) \propto 1/f$ 
relation (Papadakis \etal 2001). Although Mrk~421 (\S\ref{sec:flag}) does  
not show reliable Fourier frequency-dependent hard lags (possibly due to 
the data quality), it is still worth mentioning the evidence of similar 
dependence. Further investigations of such dependence are therefore 
essential to blazars, and to AGNs in general.

However, the mechanisms producing the X-ray emission and hard lag are rather 
different. In \mkn , synchrotron radiation of nonthermal relativistic 
electrons is responsible for the observed X-rays, and higher energy photons 
lag lower energy ones because electrons responsible for higher energy 
emission need longer time to be accelerated to the observed window. 
In contrast, in GBHCs and Seyfert galaxies, the production of the X-ray 
emission is thought to be due to thermal Comptonization of soft photons 
by hot, thermal electrons, and the hard lag originates either 
from seed soft photons (because higher energy photons have to undergo more
scatterings in the Comptonizing plasma before escape; e.g., Kazanas, Hua \& 
Titarchuk 1997; Hua, Kazanas \& Titarchuk 1997) or from the hot Comptonizing 
cloud itself (because of the changes of the energy dissipation rate; 
e.g., Poutanen \& Fabian 1999). 

The frequency-dependence of hard lags has been demonstrated to be a powerful 
tool in constraining variety of specific models responsible for the the X-ray 
emission and the associated variability in GBHCs and Seyfert galaxies 
(Papadakis et al. 2001; See also Poutanen 2001 for a review). However,  
our interpretation for the hard lag in Mrk~421 and the associated 
parameter estimates of the emitting region (
\S\ref{sec:disc:implications}) based only on the results 
from the CCF lags. Since the present work demonstrates that 
the lag between the variations at different energy bands may be 
more complicated than the existence of a fixed ``lag'', more sophisticated 
models, which will address the presence of frequency-dependent time lags, 
are needed in order to describe the variability pattern of the source at 
different energy bands. Moreover, the issue that whether the soft lags in 
TeV blazars show frequency-dependence is also crucial and worthing 
investigations. The current blazar models have not yet taken into 
account the issue of the frequency-dependence of time lags. 
Therefore, Fourier frequency-dependence of either soft or hard lag, if exist 
and confirmed, will certainly present a new constraint on the 
blazar models, specifically, on the TeV blazar models, i.e., the formation 
and structure of the shocks, particle acceleration and cooling in blazar 
jets.

		%----------------------%
\section{Conclusions} \label{sec:conc}

X-ray observations have been a powerful diagnostic for the physical
processes taking place in the vicinity of the central engines of
blazars. \sax\ observed a well-defined flare of \mkn\ on 21 April
1998, from which an X-ray hard lag was discovered in a blazar. We 
investigated with different methods the hard lag character of the 
flare, and discussed the relevant physical implications. 

We examined the conditions that may produce the hard lag variability
pattern. For the first time, the energy-dependent
acceleration time is assumed to be account for the energy-dependent
hard lags. $\xi \sim 3.06\times 10^4$ is suggested if the maximum  
characteristic energy of synchrotron emission were $\sim$ 10~keV, where \tacc\
$\sim$ \tcool . By fitting the energy-dependent hard lags using 
the energy-dependent acceleration timescale of relativistic electrons, we
found $B \sim 0.11$ Gauss, which is self-consistent with the values 
inferred from the SEDs. We then deduced that the hard lags are caused
by \tacc\ $\sim$ \tcool , and \tacc\ is energy-dependent.

Finally, we emphasize that our interpretation is based on the simplest
scenario. We consider only \tacc\ and \tcool\ without any other
complexities involved, in particular the simplified interpretation did not 
take into account the Fourier frequency-dependence of hard lags (if 
observably confirmed).  
However, to understand the relationship between electron 
acceleration/cooling and time variability, such a study should
be qualitatively meaningful, but more accurate analysis needs detailed
numerical simulations involving energy-dependent electron acceleration
timescale and information of the frequency-dependence of time lags. 
More importantly, even though there is no doubt about the sign 
of hard lag, the results presented in this work are still suggestive, 
and need further confirmation with higher quality data. Further 
investigations that whether or not the time lags are energy- and Fourier 
frequency-dependent will put important implications for the blazar 
models.
 
		%--------------------%
\section*{Acknowledgments}

The author would like to thank Annalisa Celotti and Aldo Treves for 
carefully reading the manuscript and stimulating comments. The anonymous
referee is thanked for constructive comments which led to significantly 
improve the depth and presentation of this work. This research has made
use of the standard online archive provided by the \sax\ Science Data
Center (SDC). Italian MUIR is acknowledged for financial support.

		%--------------------%

\label{lastpage}


\begin{thebibliography}{}

\bibitem[al]{} Alexander T., 1997, in Astronomical Time Series, eds. 
	D. Maoz, A. Sternberg and E.M. Leibowitz (Dordrecht:Kluwer), p.163

\bibitem[Blandford]{bld97} Blandford R., Eichler D., 1987,
	Phys. Rep., 154, 1

\bibitem[Boella et al. 1997a]{1997A&AS..122..299B} Boella G., et al.,
 	1997, A\&AS, 122, 299

\bibitem[cg]{cg99}Chiaberge M., Ghisellini G., 1999, MNRAS, 306, 551

\bibitem[Drury83]{dr83} Drury L.O'C., 1983, Rep. Prog. Phys., 46, 973

\bibitem[Edelson & Krolik 1988]{1988ApJ...333..646E} Edelson R. A., 
	Krolik J. H., 1988, ApJ, 333, 646

\bibitem[2000]{fossati00a} Fossati G. et al., 2000a, ApJ, 541, 153 

\bibitem[2000]{fossati00b} Fossati G. et al., 2000b, ApJ, 541, 166

\bibitem[1998]{gg_sed98} Ghisellini G., Celotti A., Fossati G., Maraschi
	L., Comastri A., 1998, MNRAS, 301, 451

\bibitem[gcc]{2002A&A...386..833G} Ghisellini G., Celotti A., Costamante L., 
	2002, A\&A, 386, 833
 
\bibitem[hkt]{1997ApJ...482..L57H} Hua X.M., Kazanas D., Titarchuk L., 
	1997, ApJ, 482, L57

\bibitem[Hufnagel & Bregman 1992]{1992ApJ...386..473H} Hufnagel B. R.,
	Bregman J. N., 1992, ApJ, 386, 473

\bibitem[kht]{1997ApJ...480..735K} Kazanas D., Hua X.M., Titarchuk L, 
	1997, ApJ, 480, 735

\bibitem[ktk]{ktk01}Kino M., Takahara F., Kusunose M. 2002, ApJ, 564,
	97

\bibitem[krm]{krm98} Kirk J., Rieger F., Mastichiadis A., 1998, A\&A,
	333, 452

\bibitem[lee]{2000MNRAS...318..857} Lee J.C., Fabian A.C., Reynolds 
        C.S., Brandt W.N., Iwasawa, K., MNRAS, 318, 857

\bibitem[mcp]{2000ApJ...537..L107M} Maccarone T.J., Coppi P.S., 
	Poutanen J., 2000, ApJ, L107

\bibitem[maoz]{} Maoz D., Netzer H., 1989, MNRAS, 236, 21 

\bibitem[1999]{maraschi_letter} Maraschi L. et al., 1999, ApJ, 526, L81

\bibitem[2002]{ma} Maraschi L. et al., 2002, in New visions of the X-ray 
	Universe in the XMM-Newton and Chandra Era, ed. F. Jansen, 
        in press (astro-ph/0202418)

\bibitem[mkm]{mkm88} Miyamoto S., Kitamoto S., Mitsuda K., Dotani
	T., 1988, Nat, 336, 450

\bibitem[mkk]{mi91} Miyamoto S., Kimura K., Kitamoto S., 1991, ApJ,
	383, 784
\bibitem[np]{2001ApJ...554..710} Nandra K., Papadakis I.E., 2001, 
	ApJ, 554, 710 

\bibitem[nowak]{nowak99} Nowak M.A., Vaughan B.A., Wilms J., Dove J.B., 
	Begelman M.C., 1999, ApJ, 510, 874

\bibitem[]{}Papadakis I.E., Nandra K., Kazanas D., 2001, ApJ, 554, L133

\bibitem[pl]{}Papadakis I.E., Lawrence A., 1995, MNRAS, 272, 161

\bibitem[Peterson et al. 1998]{1998PASP...110..660P} Peterson B. M., 
	Wanders I., Horne K., Collier S., Alexander T., Kaspi S., 
	Maoz D., 1998, PASP, 110, 660

\bibitem[pian]{pain98}Pian E. et al., 1998, ApJ, 492, L17 

\bibitem[pou]{} Poutanen J., 2001, Adv. Space Res., 28, 267

\bibitem[pf99]{1999MNRAS...306..L31P} Poutanen J., Fabian A.C., 1999, 
	MNRAS, 306, L31

\bibitem[smith]{1999ApJ...519..771S} Smith I.A., Liang E.P., 1999, 
	ApJ, 519, 771

\bibitem[Takahashi et al. 2000]{takahashi00} Takahashi T. et al., 2000,
	ApJ, 542, L105

\bibitem[ta]{ta01} Tanihata C., Urry C.M., Takahashi T., Kataoka J., 
 	Wagner S.J., Madejski G.M., Tashiro M., Kouda M., 
	2001, ApJ, 563, 569

\bibitem[tav]{tav01}Tavecchio F. et al., 2001, ApJ, 554, 725

\bibitem[tmg]{tmg98}Tavecchio F., Maraschi L., Ghisellini G., 1998,
	ApJ, 509, 608

\bibitem[tk95]{1995AA...300..707T} Timmer J., K\"{o}nig M., 1995, 
	A\&A, 300, 707

\bibitem[up]{up95}Urry C.M., Padovani P., 1995, PASP, 107, 803

\bibitem[wfw]{} Welsh W.F., 1999, PASP, 111, 1347

\bibitem[wp]{} White R.J., Peterson B.M., 1994, PASP, 106, 879

\bibitem[zhang]{zhang02} Zhang Y.H. et al., 2002, ApJ, 572, 762

\bibitem[zhang]{zhang99} Zhang Y.H. et al., 1999, ApJ, 527, 719

\end{thebibliography}
\end{document}